# Generation of surface soliton arrays

Yaroslav V. Kartashov, Victor A. Vysloukh,* Dumitru Mihalache,** and Lluis Torner

*ICFO-Institut de Ciencies Fotoniques, Mediterranean Technology Park, and Universitat Politecnica de Catalunya, 08860 Castelldefels (Barcelona), Spain*

We discover that at the edge of optical lattice imprinted in saturable nonlinear media one-dimensional surface solitons exist only within a band of light intensities, and that they cease to exist when the lattice depth exceeds an upper threshold. We also reveal the generation of arrays of two-dimensional surface solitons mediated by transverse modulational instability of one-dimensional solitons, a process that is found to exhibit specific features associated to properties of the optical lattice.

The presence of interfaces between linear and nonlinear materials strongly affects propagation of intense laser beams [1-3]. Such interfaces can support solitons localized at the very interface. Until recently, progress in experimental observation of surface solitons has been limited by the huge powers required for their excitation at the interfaces of natural materials. Thus, surface solitons were observed only at interfaces of linear and photorefractive media [4] and at the interface with linear layered media [5]. Recently, the concept of surface solitons in semi-infinite cubic waveguide arrays was proposed [6-8], that resulted in prediction and observation of new types of surface solitons [9,10]. All such solitons observed to date are one-dimensional. However, the technique of optical induction [11-14] allows creation of truly two-dimensional periodic interfaces. Solitons at such interfaces have never been addressed so far.

In this Letter we study the properties of solitons supported by the interface of an optical lattice with saturable nonlinearity and a linear medium. We find that in such setting one-dimensional (1D) surface solitons exist only inside a band of light intensities, and that they cease to exist when the lattice depth exceeds a critical value, the latter being a new effect arising due to the competition between nonlinearity saturation and surface effects. 1D solitons are prone to transverse modulational instabilities, previously studied for interfaces of uniform media [15-17], which may generate arrays of bulk, or two-dimensional (2D) surface solitons. Here we explore such process, for the first time to our knowledge, for solitons arrays generated at the edge of an optical lattice.

We consider the propagation of light along the $\xi$ axis of a biased photorefractive crystal. A pair of interfering plane waves induces a periodic refractive index modulation along $\eta$ axis, while along second transverse $\zeta$ axis the refractive index is uniform. An intense green-light background illumination may be used to produce a region with high photoconductivity where red-light imprinted lattice is erased, thus creating an interface between a periodic nonlinear medium and a uniform linear medium. The transition region can be made sharp by blocking the background wave in the space $\eta > 0$, so that diffraction is negligible on the scale of tenths of soliton widths. We assume that soliton beam and interface-creating waves are orthogonally polarized to each other. In this case the complex amplitude of the light field evolves according to:

$$\begin{aligned} i\frac{\partial q}{\partial \xi} &= -\frac{1}{2}\left(\frac{\partial^2 q}{\partial \eta^2} + \frac{\partial^2 q}{\partial \zeta^2}\right) - Eq\frac{|q|^2 + pR(\eta)}{1 + |q|^2 + pR(\eta)} \quad \text{at } \eta \geq 0 \\ i\frac{\partial q}{\partial \xi} &= -\frac{1}{2}\left(\frac{\partial^2 q}{\partial \eta^2} + \frac{\partial^2 q}{\partial \zeta^2}\right) \quad \text{at } \eta < 0 \end{aligned} \qquad (1)$$

Here the longitudinal $\xi$ and the transverse $\eta,\zeta$ coordinates are scaled to the diffraction length and to the beam width, respectively, the parameter $E$ describes the static biasing field applied to the crystal, $p$ is the lattice depth, and the function $R(\eta) = \sin^2(\Omega_0 \eta)$ is the profile of a lattice with frequency $\Omega_0$. We set $\Omega_0 = 2$. Our parameters correspond to beams widths $\sim 10\,\mu\text{m}$ at the wavelength $0.63\,\mu\text{m}$, launched into a SBN crystal with an electro-optic coefficient $r_{\text{eff}} = 1.8 \times 10^{-10}\,\text{m/V}$, biased with a static electric field $\sim 10^5\,\text{V/m}$. Then, $\xi = 1$ corresponds to 2 mm and $q = 1$ corresponds to peak intensities $\sim 10\,\text{mW/cm}^2$. In SBN the nonlinear coefficient for extraordinary-polarized soliton beams may exceed that for ordinary-polarized lattice beams by a factor of 20, ensuring that the lattice experiences no back-action from the soliton.

First, we address the properties of 1D solitons supported by such interfaces. We search for them in the form $q(\eta,\zeta,\xi) = w(\eta)\exp(ib\xi)$, where $w(\eta)$ is a real function and $b$ is the propagation constant, and assume that light field does not depend on $\zeta$. The quantity $U = \int_{-\infty}^{\infty} |q|^2 \, d\eta$ gives the energy flow carried by the solitons. Their profiles can be found from Eq. (1) numerically with a relaxation method. To analyze soliton stability we search for perturbed solutions of Eq. (1) in the form $q = (w + U + iV)\exp(ib\xi)$, where $U(\eta,\zeta,\xi) = u(\eta)\exp(\delta\xi)\cos(\Omega\zeta)$, $V(\eta,\zeta,\xi) = v(\eta)\exp(\delta\xi)\cos(\Omega\zeta)$, are the real and the imaginary parts of perturbations that can grow with rate $\delta$, while $\Omega$ describes the transverse modulation frequency. Substitution of these expressions into Eq. (1) and linearization around stationary solution $w$ yields the eigenvalue problem

$$\delta u = -\frac{1}{2}\left(\frac{d^2 v}{d\eta^2} - \Omega^2 v\right) + bv - Ev\frac{w^2 + pR}{1 + w^2 + pR},$$
$$\delta v = \frac{1}{2}\left(\frac{d^2 u}{d\eta^2} - \Omega^2 u\right) - bu + Eu\left[\frac{3w^2 + pR}{1 + w^2 + pR} - \frac{2w^4 + 2pRw^2}{(1 + w^2 + pR)^2}\right], \quad (2)$$

for $\eta \geq 0$, while at $\eta < 0$ all terms with $E$ should be omitted.

There are two simplest types of surface solitons: odd (centered in the nearest to interface lattice channel) and even (centered between the first and second lattice channels) as shown in Fig. 1. The dependence of energy flow on propagation constant for odd and even solitons is depicted in Fig. 2(a). In the low-energy limit, surface solitons acquire pronounced shape modulations and tend to penetrate deep into the lattice region (Figs. 1(a),1(b)). Such solitons exist only above a power threshold, similarly to surface solitons in cubic lattices. With increase of $U$ (or $b$) the center of odd soliton gradually shifts toward the first lattice minimum, while soliton itself becomes well-localized (Figs. 1(c) and 1(d)). Thus, the interface effectively *repels* high-energy solitons. This is in contrast to surfaces in cubic media, where energy concentrates in first lattice channel as $b \to \infty$. Due to such repulsion, surface solitons can exist in the first lattice channel only below a threshold energy flow. In the presence of saturation surface solitons exist inside a band of energy flows or propagation constants (see Fig. 2(a)). In the lower $b_{\text{low}}$ and upper $b_{\text{upp}}$ cutoffs profiles of odd and even solitons coincide. We discovered that surface solitons at saturable interfaces cease to exist when $p$ exceeds a critical value $p_{\text{cr}}$ (Fig. 2(b)). Thus, at $E = 10$ one has $p_{\text{cr}} \approx 4.01$. A similar behavior was encountered for other values of the biasing static field $E$. The existence domain is found to shrink at $E = E_{\text{cr}}$ (Fig. 2(c)), while with increase of $E$ the range of accessible energy flows of surface waves increases. We found a qualitatively similar behavior for other types of saturable interfaces, indicating a universality of this phenomenon.

Linear stability analysis indicates that odd surface solitons are stable with respect to one-dimensional perturbations ($\Omega = 0$), except for narrow regions near lower and upper cutoffs where $dU/db \leq 0$ (Fig. 2(d)). At the same time, even solitons happen to be unstable in the entire domain of their existence. The perturbation growth rates for odd and even surface solitons coincide at the boundaries of existence domain.

The central issue addressed in this Letter is the evolution of the 1D solitons when higher-dimensional dynamics along the transverse coordinate $\zeta$ is included. In this case, development of transverse modulational instability (TMI) of surface wave stripe along $\zeta$ may result in formation of 2D soliton arrays. To elucidate the specific features of TMI at interfaces with lattice, we calculated the growth rates for odd and even solitons at modulation frequencies $\Omega > 0$ (Fig. 3). All instabilities that were encountered are of exponential type. Growth rate is nonzero only within a finite frequency band: $\delta$ vanishes for odd solitons and remains nonzero for even solitons when $\Omega \to 0$.

Direct integration of Eq. (1) enables to determine which structures emerge in the developed stage of TMI. We solved Eq. (1) with the input conditions $q|_{\xi=0} = w + (u + iv)\cos(\Omega\zeta)$, where $u, v \ll w$ are obtained from Eq. (2). Thus, input field distributions are bright near-surface stripes localized along $\eta$ axis, and are only slightly modulated along $\zeta$ axis. Three typical regimes were encountered. For small modulation frequencies, when energy concentrated within one filament with scale $2\pi/\Omega$ (along $\zeta$ axis) substantially exceeds energy necessary for excitation of 2D solitons, the 1D surface waves break into many fragments (Fig. 4(a)), that either move away of the surface or form sets of 2D surface waves. The specific feature of TMI at the interface is that decay may be accompanied by strong radiation emission into the lattice, but almost no radiation penetrates into the linear medium. Importantly, the interface repels beams with too high or too low energy flows, consistent with limited existence domain in $U$ for surface solitons and only filaments with proper energy remain in the vicinity of interface.

A second important regime is encountered when energy flow within one filament is of the same order as the energy required for formation of a single 2D surface soliton. In this case, after some transient regime accompanied by radiation (Fig. 4(b)) an array of localized 2D surface solitons is formed. Under suitable conditions, this regime occurs for modulation frequency yielding the maximal growth rate. A third regime occurs when energy concentrated within one filament is not high enough for formation of 2D surface wave. This typically happens close to the high-frequency edge of the instability domain. Then, a shallow modulation of surface wave occurs, without wave break-up (Fig. 4(c)). Simulations show that arrays of well-localized 2D surface waves emerging due to TMI of 1D waves are robust and may propagate undistorted over huge distances (e.g., the array depicted in Fig. 4(b) keeps its structure for more than $10^4$ propagation units, exceeding any experimentally feasible crystal lengths by several orders of magnitude). By tuning the energy flow of input quasi-1D waves and frequency of transverse modulation one may generate arrays with various separations between neighboring spots along $\zeta$ axis.

We thus conclude by stressing that we introduced the properties of solitons attached at the surface of a lattice imprinted in media with saturable nonlinearity. The existence conditions of such solitons exhibit new threshold effects versus energy flow and lattice depth. We showed how transverse modulational instabilities of one-dimensional solitons lead to the generation of robust two-dimensional surface solitons arrays.

*Also with Universidad de las Americas, Puebla, Mexico. **Also with Institute of Atomic Physics, Bucharest, Romania.

# References with titles

# References without titles

# Figure captions

Figure 1. Profiles of odd (a),(c) and even (b),(d) surface solitons. In (a) and (b) $b=5.2$. In (c) and (d) $b=7$. Gray regions correspond to $R(\eta)>1/2$, while in white regions $R(\eta)\leq 1/2$. In all cases $E=10$ and $p=2$.

Figure 2 (color online). (a) Energy flow vs propagation constant for odd (black line) and even (red line) surface solitons. Domains of existence of surface solitons on $(p,b)$ plane at $E=10$ (b) and on $(E,b)$ plane at $p=2$ (c). (d) Perturbation growth rate vs propagation constant for odd (black line) and even (red line) solitons at $\Omega=0$, $E=10$, $p=2$. Circles indicate the points where odd and even soliton profiles and perturbation growth rates coincide.

Figure 3. Perturbation growth rate for odd (a) and even (b) surface solitons vs propagation constant $b$ and modulation frequency $\Omega$ at $E=10$ and $p=2$.

Figure 4 (color online). Development of transverse modulational instability of quasi-1D surface wave at $\Omega=0.6$ (a), 1.3 (b), and 1.7 (c). Field modulus distributions are shown at different distances. White dashed lines indicate interface position. Quasi-1D surface wave corresponds to $b=5.5$, $E=10$ and $p=2$.

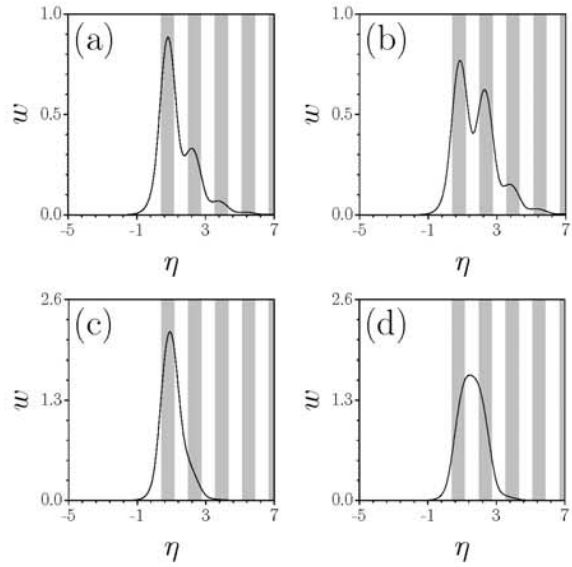

Figure 1. Profiles of odd (a),(c) and even (b),(d) surface solitons. In (a) and (b) $b = 5.2$. In (c) and (d) $b = 7$. Gray regions correspond to $R(\eta) > 1/2$, while in white regions $R(\eta) \leq 1/2$. In all cases $E = 10$ and $p = 2$.

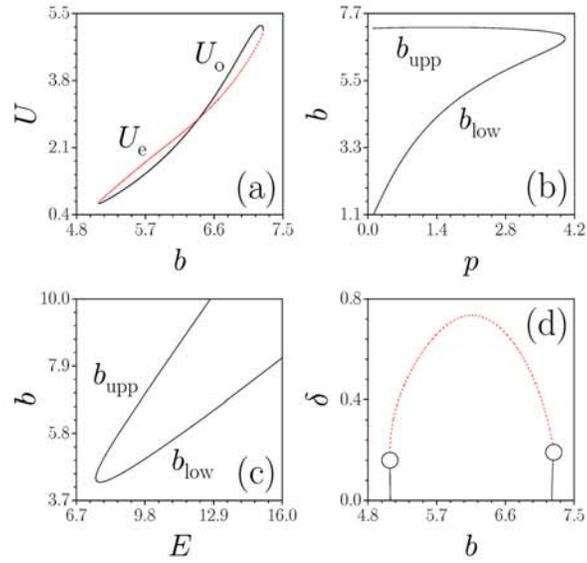

Figure 2 (color online). (a) Energy flow vs propagation constant for odd (black line) and even (red line) surface solitons. Domains of existence of surface solitons on $(p,b)$ plane at $E=10$ (b) and on $(E,b)$ plane at $p=2$ (c). (d) Perturbation growth rate vs propagation constant for odd (black line) and even (red line) solitons at $\Omega=0$, $E=10$, $p=2$. Circles indicate the points where odd and even soliton profiles and perturbation growth rates coincide.

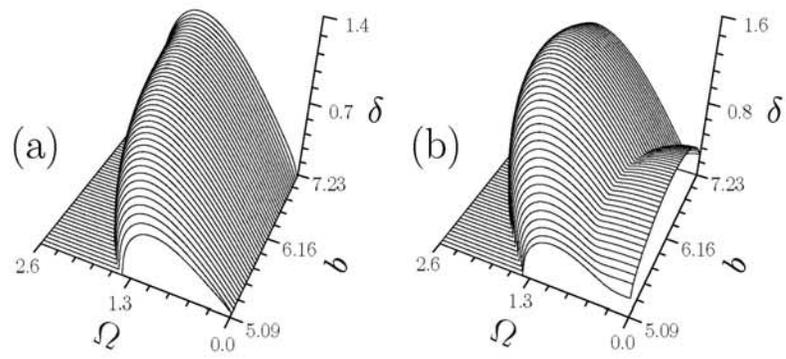

Figure 3. Perturbation growth rate for odd (a) and even (b) surface solitons vs propagation constant $b$ and modulation frequency $\Omega$ at $E = 10$ and $p = 2$.

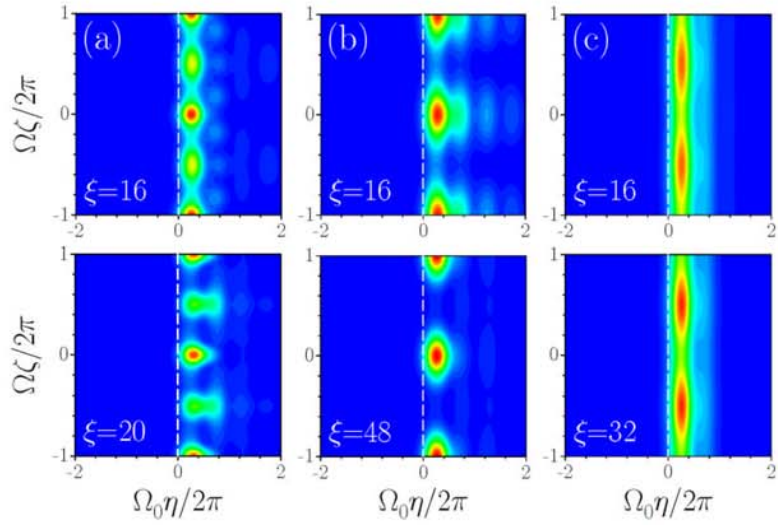

Figure 4 (color online).   Development of transverse modulational instability of quasi-1D surface wave at $\Omega = 0.6$ (a), 1.3 (b), and 1.7 (c). Field modulus distributions are shown at different distances. White dashed lines indicate interface position. Quasi-1D surface wave corresponds to $b = 5.5$, $E = 10$ and $p = 2$.